%% file: main.tex
\documentclass{article}
\usepackage{spconf}
	
\usepackage[dvipsnames]{xcolor}
\usepackage{times}
\usepackage{epsfig}
\usepackage{graphicx}
\usepackage{amsmath}
\usepackage{amssymb}
\usepackage{multirow}
\usepackage{collcell}
\usepackage{tabularx}
\usepackage{diagbox}
\usepackage{array}
\usepackage{color, colortbl}
\usepackage{subcaption}
\usepackage{stackengine}
\usepackage{tikz}
\usepackage{dblfloatfix}
\usepackage{xcolor}
\usepackage{url}

\usepackage{verbatim}
\usepackage{balance}

\usepackage[normalem]{ulem}

\newcommand{\new}[1]{\uline{#1\/}}

\makeatletter
\newcommand{\multiline}[1]{%
  \begin{tabularx}{\dimexpr\linewidth-\ALG@thistlm}[t]{@{}X@{}}
    #1
  \end{tabularx}
}
\makeatletter
\def\hlinewd#1{%
  \noalign{\ifnum0=`}\fi\hrule \@height #1 \futurelet
  \reserved@a\@xhline}
  
\graphicspath{ {./images} }

\title{HistoKT: Cross Knowledge Transfer in Computational Pathology
\vspace{-4mm} }
\name{\begin{tabular}{c}Ryan Zhang$^\textit{1*}$\thanks{*Equal contribution}, Jiadai Zhu$^\textit{1*}$, Stephen Yang$^\textit{1*}$, Mahdi S. Hosseini$^\textit{2*}$, Angelo Genovese$^\textit{3}$, \\ Lina Chen$^\textit{4}$, Corwyn Rowsell$^\textit{5}$, Savvas Damaskinos$^\textit{6}$, Sonal Varma$^\textit{7}$, Konstantinos N. Plataniotis$^\textit{1}$\end{tabular}\vspace{-4mm}}
\address{%
\small $^\textit{1}$University of Toronto, Canada $ $
$^\textit{2}$University of New Brunswick, Canada  \vspace{-1mm} \\
\small $^\textit{3}$Universit\`a degli Studi di Milano, Italy $ $
$^\textit{4}$Sunnybrook Health Sciences Centre, Canada \vspace{-1mm} \\
\small $^\textit{5}$St. Michaels Hospital, Canada $ $
$^\textit{6}$Huron Digital Pathology, Canada $ $
$^\textit{7}$Kingston Health Sciences Center, Canada
\\
\color{purple}{\url{https://github.com/mahdihosseini/HistoKT}}
\vspace{-3mm}
}

\begin{document}
\ninept

\maketitle

\begin{abstract}
\vspace{-1mm}
The lack of well-annotated datasets in computational pathology (CPath) obstructs the application of deep learning techniques for classifying medical images. 
Many CPath workflows involve transferring learned knowledge between various image domains through transfer learning. Currently, most transfer learning research follows a model-centric approach, tuning network parameters to improve transfer results over few datasets. In this paper, we take a data-centric approach to the transfer learning problem and examine the existence of generalizable knowledge between histopathological datasets. First, we create a standardization workflow for aggregating existing histopathological data. We then measure inter-domain knowledge by training ResNet18 models across multiple histopathological datasets, and cross-transferring between them to determine the quantity and quality of innate shared knowledge. Additionally, we use weight distillation to share knowledge between models without additional training. We find that hard to learn, multi-class datasets benefit most from pretraining, and a two stage learning framework incorporating a large source domain such as ImageNet allows for better utilization of smaller datasets. Furthermore, we find that weight distillation enables models trained on purely histopathological features to outperform models using external natural image data.

\end{abstract}



\input{Intro}

\input{Methods}

\input{Results}
\input{conclusion}

{
\small
\balance
\bibliographystyle{IEEEbib}
\vspace{-3mm}
\bibliography{egbib}
}

\end{document}

%% file: Intro.tex
\vspace{-4mm}
\section{Introduction}
\vspace{-2mm}


Currently in the United States, there are a reported $3.94$ pathologists per $100,000$ people. In Canada, this number rises slightly to $4.81$ pathologists per $100,000$ people \cite{wherearethepathologists}. This severe scarcity of pathologists, combined with a rigorous set of duties that involves patient care and extraneous specimen diagnoses, results in decreased diagnosis quality and diminished patient experience \cite{wherearethepathologists}. To alleviate these burdens, the computational pathology (CPath) field has created numerous computer-aided diagnosis (CAD) tools to assist pathologist diagnoses \cite{dlinmedicalimaging}. These CAD tools utilize computer vision techniques and neural network architectures to solve a plethora of tasks, including classification, segmentation, and localization \cite{dlinmedicalimaging}.

However, challenges in implementing CAD systems arise in part due to limitations in pathology datasets. Namely, despite the presence of large pathology datasets, the lack of proper annotations or labels hinders development of supervised neural networks \cite{dlinmedicalimaging}. Additionally, different standards for staining histology slides and varying optical configurations introduce further complications when creating CAD systems \cite{StainInvariantFeatures,AIAndCPath,QuantifyingDataAugAndStainNormForCPath}. Combined, these factors result in a valuable data landscape comprised of sparse and non-comprehensive datasets.

Transfer learning is widely used in machine learning to compensate for the absence of comprehensive annotated datasets. Using transfer learning, knowledge gained from one source domain can be applied to problems on another target domain. In CPath, transfer learning using either natural image datasets or other pathology datasets enables networks to generalize to specific target domains where labeled data is scarce \cite{DoubleTransferLearning,StepwiseTransferLearning,ADPtransfer, TumorClassificationTransferLearning}.


The merits of transfer learning in CPath are clear: models achieve higher metrics on a target domain when pretrained on a relevant source domain dataset \cite{DoubleTransferLearning,StepwiseTransferLearning,ADPtransfer, TumorClassificationTransferLearning}. These benefits are explored in previous works through model-centric approaches. While these approaches are useful in exploring quality in various model architectures, they ignore key issues introduced when these models are applied to other datasets. Due to a lack of standardized data preparation when transferring knowledge between datasets, models trained on two distinct datasets are likely to learn at two distinct biological scales. Previous work has shown that models trained from a dataset gathered by one pathology laboratory can underperform when applied to images gathered by a separate laboratory \cite{QuantifyingDataAugAndStainNormForCPath}. Furthermore, dataset choice for the source domain is not well explored, with little known about what makes a ``good'' dataset. These issues can only be resolved through a data-centric approach towards exploring the interactions of source domain data with target domain data under transfer learning.


In this paper we introduce the following contributions: 
\textit{i)}~we propose a standardized platform to aggregate learned histopathological knowledge, including an image standardization workflow, training, and a tuning pipeline
; 
\textit{ii)}~we examine the potential for aggregation of learned knowledge between multiple pathological datasets. Using cross transfer over nine classification task datasets, we evaluate both the quantity and quality of transferable information between datasets; 
\textit{iii)}~we propose weight distillation: a method for combining learned information from encoders trained on separate datasets. 
%
\textit{iv)}~we assess the utility of large natural image domains (ImageNet) as a source domain with two stage transfer learning;
\textit{v)}~we visualize the transferred knowledge using t-SNE plots and Grad-CAM images.

%% file: Methods.tex
\graphicspath{ {./images/main/} }

\input{DatasetTable}

\input{Workflow}

\vspace{-3mm}
\section{Methods}
\vspace{-2mm}

We introduce our pipeline for knowledge transfer, which includes dataset preprocessing, model training, and evaluation. 
In our evaluation, we considered the databases summarized in Tab.~\ref{tab:DatasetTable}.
The overall HistoKT workflow is summarized in Fig.~\ref{fig:workflow}.

\vspace{-2mm}
\subsection{Preprocessing}
\vspace{-2mm}

Datasets were standardized according to our pipeline, consisting of rescaling, cropping, and reflection wrapping to match the benchmark dataset, ADP. ADP was chosen as a benchmark due to its coverage of various histological tissue types. 

Each image in a given dataset was rescaled to the common pixel resolution of 1 $\mu m$ using the \texttt{scikit-image} library. If the resultant image is larger than $272$ pixels in either dimension, the image is cropped into $272 \times 272$ patches, with $50$\% overlap in either direction. If the rescaled image is smaller than $272$ pixels in either dimension, the image is reflection wrapped. After cropping, background images were filtered; images that had low contrast, with pixels falling between the $5$th percentile and $99$th percentile having less than $5$\% coverage of the colour span ($0-255$), were removed.

This pipeline was chosen for maintaining biological scale across datasets, so that models trained on one dataset operate at the same scale when applied to another dataset. 



\vspace{-3mm}
\subsection{Training}
\vspace{-2mm}

ResNet18 \cite{ResNet} was chosen as the baseline model due to its wide use throughout literature, as well as its relatively small number of parameters compared to other commonly selected networks. 

For all experiments, training was conducted using \texttt{PyTorch}, utilizing NVIDIA Tesla V100 Tensor Core GPUs. For all baseline results, models were trained from random initialization on a given dataset, using the RMSGD optimizer \cite{adasPaper}, with an initial learning rate of $0.03$, momentum of $0.9$, weight decay of $5e-4$, and all other parameters left as default \cite{SGDmomentum}. Multi-labeled datasets, ADP and BCSS, are trained with a weighted one-vs-all cross entropy loss, while all other datasets are trained with cross entropy loss. Baseline models were trained for $250$ epochs, with three trials for each dataset. The model with the highest validation accuracy for a given epoch was taken as the baseline weight for further evaluation.


\vspace{-2mm}
\subsection{Tuning} 
\vspace{-2mm}

Three primary methods were tested for tuning on a target domain: no tuning, fine-tuning, and deep-tuning. For no tuning, we take the encoder trained on a source domain and evaluate the encoder on the same domain. We denote fine-tuning to be tuning with all layers frozen except the final fully connected (FC) layer, and deep-tuning as tuning where no layers are frozen.

Our tuning procedure uses the AdamP optimizer \cite{AdamP}, with weight decay set to $5e-4$, and all other parameters left as default. Learning rates were determined through a grid search. 
A learning rate scheduler was used which reduced the learning rate by a factor of two every $20$ epochs. Models were trained for $250$ epochs, and three trials were run for every learning rate and target domain combination.


\vspace{-4mm}
\subsection{Transferability}
\vspace{-2mm}

Transferability is evaluated using comparison matrices, as shown in Tab.~\ref{tab:confusion_matrix} and Tab.~\ref{tab:norm_ImageNet_confusion_matrix}. Along the diagonals are the average top-1 test accuracies for each dataset trained from random initialization using the methodology described in the training section. We then pick the best baseline models with respect to top-1 test accuracy as the candidate model to perform tuning, and present the average test accuracy on the target datasets. All of the (off-diagonal) results are deep-tuned, as deep-tuning greatly outperforms fine-tuning in most datasets, as shown in Tab.~\ref{tab:deepVSfine}. Deep-tuning also allows us to compare the difference in learned representations with t-SNE plots, as fine-tuning does not change the encoder weights. 

\input{confusion_matrix}
\input{norm_ImageNet_confusion_matrix}
\input{deepVSfineTable}

\vspace{-1mm}
\subsection{Weight Distillation}
\vspace{-1mm}

Taking the baseline weights from top performing models, we perform weight distillation as a secondary method for evaluating the potential for knowledge aggregation. Since all of our models use the same architecture\textemdash ResNet18\textemdash each model differs only by the dataset it is trained on. For each layer of the model, we unfold the 4-D weight tensors $\mathbf{W}^l _{dataset} \in {\mathbb{R}} ^{w \times h \times n_i \times n_o}$, where $l$ is the layer number, into 2-D weight tensors $\mathbf{\overline{W}} ^l _{dataset} \in {\mathbb{R}} ^{n_o \times k}$, $k = w \times h \times n_i$. To combine weights from models trained on different datasets, we stack the unfolded weight tensors from all source datasets on top of each other to create a new weight tensor. For example, a new weight tensor for the $5^{th}$ convolutional layer \vspace{-2mm}
$$ \mathbf{\overline{W}} ^5 _{C} = \left[ \mathbf{\overline{W}} ^5 _{ADP},  \mathbf{\overline{W}} ^5 _{CRC} \right]^T \vspace{-2mm} $$ 
is created using $\mathbf{\overline{W}} ^5 _{ADP}$ from layer five of a model trained from random initialization on ADP and $\mathbf{\overline{W}} ^5 _{CRC}$ from layer five of a model trained from random initialization on CRC. We then apply Singular Value Decomposition (SVD) \vspace{-2mm}
$$ \mathbf{\overline{W}} ^l _{C} = \mathbf{\overline{U}} ^l _{C} \mathbf{\overline{\Lambda}} ^l _{C}  \mathbf{{\overline{V}} ^l _{C} }^T \vspace{-2mm} $$
to create factorized matrices. We take the first $n_o$ rows of $\mathbf{\overline{\Lambda}} ^l _{C}$ and the first $n_o$ rows and $n_o$ columns of $\mathbf{\overline{U}} ^l _{C}$ to keep only the most important filter values in the combined weight tensor, i.e. ${\overline{\mathbf{\Lambda}^l}}' = \overline{\mathbf{\Lambda}} ^l _{C}[:n_o, :]$ and $\overline{\mathbf{U}^l}' = \overline{\mathbf{U}} ^l _{C}[:n_o, :n_o]$. In this way, $\overline{\mathbf{W}}^{l'} = \mathbf{\overline{U}}^{l'} \overline{\mathbf{\Lambda}}^{l'} {\overline{\mathbf{V}}^l_C}^T \in \mathbb{R}^{n_o \times k}$. Then, we fold ${\mathbf{\overline{W}}^{l}}'$ back to a 4-D tensor ${\mathbf{W}^l}' \in \mathbb{R} ^{w \times h \times n_i \times n_o}$. For non-convolutional 1-D layers, which include batch normalization and linear layers, the resultant vector $\mathbf{w}^{l'} \in \mathbb{R}^n$ is the mean of the corresponding vectors in all input models. For 2-D linear weights, the same SVD process is carried out. The resulting model is deep tuned according to our tuning methodology.

\vspace{-2mm}
\subsection{Evaluation}
\vspace{-2mm}

We train our models on the training set and use the validation set to select the best performing model for each $250$ epoch trial. Based on the validation accuracy, a single best performing model is selected to be evaluated on the test set. All results reported in this paper are test set metrics averaged over three runs. To evaluate the model on the test set, we calculate the test accuracy.
All results are summarized in the results section.

Moreover, we use t-Distributed Stochastic Neighbor Embedding (t-SNE) \cite{t-SNE} and Grad-CAM \cite{Grad-CAMpaper} to visualize our results. t-SNE visualizes high-dimensional data by giving each datapoint a location in a 2-D map. We also use Grad-CAM to visualize activation heatmaps of the network on various classes, where a warmer colour intensity corresponds to the amount of influence an image region has on a model prediction. 

%% file: DatasetTable.tex
\newcommand{\centered}[1]{\begin{tabular}{l} #1 \end{tabular}}

\begin{table*}[h]
    \setlength\tabcolsep{1pt} 
    \center
    \caption{Dataset Information}\vspace{-3mm}
	\label{tab:DatasetTable}
    \scriptsize{
    \begin{tabular}{ c||cc|c|p{3cm}|c|c|c|c||c|c|c}
        \hlinewd{1pt}
        \multicolumn{9}{c||}{\textbf{Original Dataset Information}} & \multicolumn{3}{c}{\textbf{Number of Extracted Patches}} \\
        \hlinewd{1pt}
        \multirow{2}{*}{\textbf{Dataset Name}} & \multirow{2}{*}{\textbf{Tissue Type}} & \multirow{2}{*}{\textbf{Diagnostic}} & \multirow{2}{*}{\textbf{Staining}} & \multirow{2}{*}{\textbf{Scanner}} & \multirow{2}{*}{\textbf{Classes}} & \multirow{2}{*}{\textbf{Dataset Size}} & \multirow{2}{*}{\textbf{Image Size}} & \textbf{Pixel} & \textbf{Training} & \textbf{Validation} & \textbf{Test}\\
        &&&&&&&& \textbf{Resolution} & \textbf{Images} & \textbf{Images}& \textbf{Images}\\
        \hline
        \hline
        ADP~\cite{ADP} & Multi-organ & Histology (healthy) & H\&E & Huron TissueScope LE1.2 & 33 & 17668 & 272 $\times$ 272 & 1 $\mu m$ & 14134 & 1767 & 1767 \\
        
        MHIST~\cite{MHISTpaper} & Colorectal polyps & Cancer & H\&E & Aperio AT2 & 2 & 3152 & 224 $\times$ 224 & 1.25 $\mu m$ & 1740 & 435 & 977 \\
        
        BACH~\cite{BACHdataset} & Breast & Cancer & H\&E & Leica ICC50 HD & 4 & 400 & 2048 $\times$ 1536 & 0.42 $\mu m$ & 958 & 240 & 1199 \\
        
        AJ-Lymph~\cite{AJ-LymphPaper} & Lymph nodes & Lymphoma & H\&E & N/A & 3 & 374 & 1388 $\times$ 1040 & 0.25 $\mu m$ & 299 & 37 & 38 \\
        
        PCam~\cite{PCamPaper} & Lymph nodes & Lymphoma & H\&E & Pannoramic 250 Flash II, NanoZoomer-XR Digital slide scanner C12000-01 & 2 & 294912 & 96 $\times$ 96 & 0.972 $\mu m$ & 2000 & 400 & 32322 \\
        
        CRC~\cite{CRCpaper} & Colon \& rectum & Histopathology & H\&E & Online & 7 & 107000 & 224 $\times$ 224 & 0.5 $\mu m$ & 14000 & 1750 & 1750 \\
        
        GlaS~\cite{GlaSdataset} & Intestinal glands & Cancer & H\&E & Zeiss MIRAX MIDI Slide Scanner & 2 & 165 & Various & 0.62 $\mu m$ & 163 & 112 & 40 \\
        
        OS~\cite{TCIA} & Bone & Osteosarcoma & H\&E & N/A & 3 & 1144 & 1024 $\times$ 1024 & 1 $\mu m$ & 13227 & 1653 & 1654 \\
        
        BCSS~\cite{BCSSpaper} & Breast & Histopathology & H\&E & Online & 10 & 151 & Various & 0.25 $\mu m$ & 14288 & 1786 & 1787 \\
        \hlinewd{1pt}
    \end{tabular}
    }
\vspace{-4mm}
\end{table*}

%% file: Workflow.tex
\graphicspath{ {./images/} }

\begin{figure*}[t]
\centering
\includegraphics[width=0.85\linewidth]{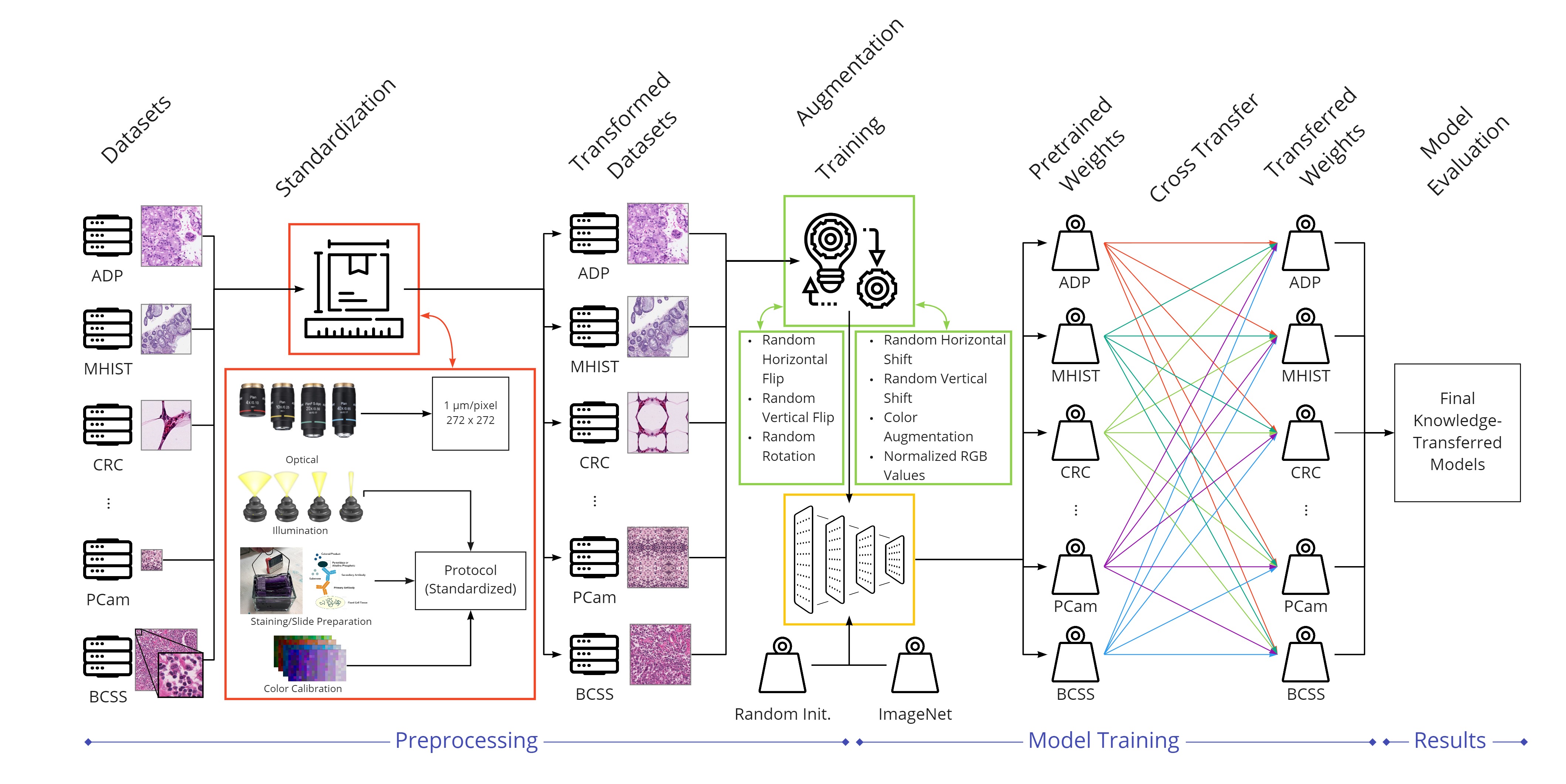}
\vspace{-5mm}
\caption{In order to use a new dataset as a source domain for transfer learning, it must first be preprocessed via standardization to the ADP dataset format. Afterwards, model training and tuning enables knowledge cross transfer to a task-specific target domain. This cross transfer is carried out as tuning of a source domain trained encoder. We tune all datasets on all source datasets to evaluate performance.}
\label{fig:workflow}
\vspace{-3mm}
\end{figure*}

%% file: confusion_matrix.tex
\newcolumntype{P}[1]{>{\centering\arraybackslash}p{#1}}
\newcolumntype{M}[1]{>{\centering\arraybackslash}m{#1}}

\definecolor{ashgrey}{rgb}{0.7, 0.75, 0.71}
\newcommand{\baseline}{\cellcolor{ashgrey}}

\definecolor{green1}{HTML}{63BE7B}
\definecolor{green2}{HTML}{77C68C}
\definecolor{green3}{HTML}{8ACE9D}
\definecolor{green4}{HTML}{9DD6AD}
\definecolor{green5}{HTML}{B0DDBD}
\definecolor{green6}{HTML}{C3E4CE}
\definecolor{green7}{HTML}{D6EDDE}
\definecolor{green8}{HTML}{E9F5EF}
\newcommand{\highone}{\cellcolor{green1}}
\newcommand{\hightwo}{\cellcolor{green2}}
\newcommand{\highthree}{\cellcolor{green3}}
\newcommand{\highfour}{\cellcolor{green4}}
\newcommand{\highfive}{\cellcolor{green5}}
\newcommand{\highsix}{\cellcolor{green6}}
\newcommand{\highseven}{\cellcolor{green7}}
\newcommand{\higheight}{\cellcolor{green8}}

\definecolor{red1}{HTML}{F8696B}
\definecolor{red2}{HTML}{F87B7D}
\definecolor{red3}{HTML}{F98D90}
\definecolor{red4}{HTML}{F99FA2}
\definecolor{red5}{HTML}{F9B2B5}
\definecolor{red6}{HTML}{FAC4C7}
\definecolor{red7}{HTML}{FBD7DA}
\definecolor{red8}{HTML}{FBE9EC}
\newcommand{\lowone}{\cellcolor{red1}}
\newcommand{\lowtwo}{\cellcolor{red2}}
\newcommand{\lowthree}{\cellcolor{red3}}
\newcommand{\lowfour}{\cellcolor{red4}}
\newcommand{\lowfive}{\cellcolor{red5}}
\newcommand{\lowsix}{\cellcolor{red6}}
\newcommand{\lowseven}{\cellcolor{red7}}
\newcommand{\loweight}{\cellcolor{red8}}

\newcommand\cellRedBG[1]{\cellcolor{red!#1!white}}
\newcommand\cellGreenBG[1]{\cellcolor{green1!#1!white}}

\begin{table*}[t]
    \setlength\tabcolsep{1pt} 
    \center
    \caption{HistoKT Matrix (Top-1 Test Accuracy)}
    \vspace{-2mm}
	\label{tab:confusion_matrix}
    \tiny{
    \begin{tabular}{c|P{1.6cm} P{1.6cm} P{1.6cm} P{1.6cm} P{1.6cm} P{1.6cm} P{1.6cm} P{1.6cm} P{1.6cm}}
    \hlinewd{1pt}
        \multirow{2}{*}{\backslashbox{\textbf{Source}}{\textbf{Target}}} 
        & \multirow{2}{*}{\textbf{ADP}}
        & \multirow{2}{*}{\textbf{MHIST}}
        & \multirow{2}{*}{\textbf{BACH}}
        & \multirow{2}{*}{\textbf{AJ-LYMPH}}
        & \multirow{2}{*}{\textbf{PCam}}
        & \multirow{2}{*}{\textbf{CRC}}
        & \multirow{2}{*}{\textbf{GlaS}}
        & \multirow{2}{*}{\textbf{OS}}
        & \multirow{2}{*}{\textbf{BCSS}}
        \\
        & \multirow{2}{*}{} & \multirow{2}{*}{} & \multirow{2}{*}{}& \multirow{2}{*}{} & \multirow{2}{*}{} & \multirow{2}{*}{} & \multirow{2}{*}{} & \multirow{2}{*}{}
        \\
        \hline
\textbf{ADP} & {93.56$_{0.4}$} & \cellRedBG{27.71}{78.74$_{1.39}$} & \cellGreenBG{77.31}{93.11$_{0.05}$} & \cellGreenBG{78.18}{94.74$_{2.63}$} & \cellRedBG{55.0}{76.37$_{0.92}$} & \cellGreenBG{100.0}{99.3$_{0.12}$} & \cellGreenBG{55.56}{88.33$_{7.22}$} & \cellGreenBG{100.0}{94.92$_{0.53}$} & \cellRedBG{26.07}{97.57$_{0.05}$}\\
\textbf{MHIST} & \cellRedBG{45.5}{93.35$_{0.07}$} & {80.83$_{1.41}$} & \cellRedBG{40.7}{84.35$_{0.91}$} & \cellGreenBG{49.09}{91.23$_{4.02}$} & \cellRedBG{25.96}{77.68$_{0.61}$} & \cellRedBG{21.25}{98.93$_{0.12}$} & \cellRedBG{32.5}{80.83$_{3.82}$} & \cellRedBG{43.44}{93.11$_{0.37}$} & \cellRedBG{30.16}{97.55$_{0.04}$}\\
\textbf{BACH} & \cellRedBG{16.68}{93.52$_{0.03}$} & \cellRedBG{39.61}{77.35$_{0.43}$} & {90.44$_{0.85}$} & \cellGreenBG{41.82}{90.35$_{5.48}$} & \cellGreenBG{100.0}{79.36$_{0.66}$} & \cellRedBG{34.75}{98.82$_{0.07}$} & \cellRedBG{50.5}{77.5$_{10.9}$} & \cellRedBG{31.05}{93.71$_{0.31}$} & \cellRedBG{27.82}{97.56$_{0.09}$}\\
\textbf{AJ} & \cellRedBG{41.75}{93.37$_{0.02}$} & \cellRedBG{55.0}{75.54$_{0.99}$} & \cellRedBG{55.0}{81.51$_{0.19}$} & {87.72$_{4.02}$} & \cellRedBG{11.46}{78.34$_{0.9}$} & \cellRedBG{55.0}{98.65$_{0.17}$} & \cellRedBG{55.0}{76.67$_{6.29}$} & \cellRedBG{50.46}{92.76$_{0.36}$} & \cellRedBG{55.0}{97.39$_{0.07}$}\\
\textbf{PCam} & \cellGreenBG{26.09}{93.62$_{0.04}$} & \cellRedBG{49.19}{76.22$_{1.32}$} & \cellRedBG{29.63}{86.54$_{1.21}$} & \cellRedBG{55.0}{85.96$_{1.52}$} & {78.4$_{1.39}$} & \cellRedBG{52.75}{98.67$_{0.26}$} & \cellRedBG{23.5}{82.5$_{9.01}$} & \cellRedBG{55.0}{92.54$_{0.21}$} & \cellRedBG{37.47}{97.5$_{0.03}$}\\
\textbf{CRC} & \cellGreenBG{94.62}{94.22$_{0.04}$} & \cellRedBG{18.71}{79.8$_{1.13}$} & \cellGreenBG{100.0}{94.16$_{0.58}$} & \cellGreenBG{100.0}{97.37$_{0.0}$} & \cellRedBG{10.27}{78.39$_{2.34}$} & {99.03$_{0.06}$} & \cellGreenBG{100.0}{92.5$_{0.0}$} & \cellRedBG{13.3}{94.58$_{0.07}$} & \cellRedBG{47.1}{97.44$_{0.06}$}\\
\textbf{GlaS} & \cellRedBG{55.0}{93.29$_{0.09}$} & \cellRedBG{28.29}{78.68$_{1.59}$} & \cellRedBG{48.27}{82.85$_{1.26}$} & \cellRedBG{32.5}{86.84$_{2.63}$} & \cellRedBG{48.63}{76.66$_{0.95}$} & \cellRedBG{25.75}{98.9$_{0.26}$} & {85.0$_{5.0}$} & \cellRedBG{44.27}{93.07$_{0.65}$} & \cellRedBG{40.39}{97.48$_{0.04}$}\\
\textbf{OS} & \cellGreenBG{100.0}{94.27$_{0.12}$} & \cellRedBG{34.1}{77.99$_{0.74}$} & \cellGreenBG{90.45}{93.72$_{0.24}$} & \cellGreenBG{70.91}{93.86$_{3.04}$} & \cellRedBG{26.23}{77.67$_{2.11}$} & \cellGreenBG{71.43}{99.2$_{0.23}$} & \cellGreenBG{82.22}{90.83$_{2.89}$} & {94.74$_{0.3}$} & \cellRedBG{32.49}{97.53$_{0.02}$}\\
\textbf{BCSS} & \cellGreenBG{72.92}{94.03$_{0.01}$} & \cellGreenBG{100.0}{81.88$_{0.67}$} & \cellGreenBG{88.06}{93.61$_{0.42}$} & \cellGreenBG{70.91}{93.86$_{4.02}$} & \cellGreenBG{30.29}{78.52$_{0.27}$} & \cellGreenBG{25.72}{99.05$_{0.14}$} & \cellGreenBG{73.33}{90.0$_{4.33}$} & \cellRedBG{13.72}{94.56$_{0.06}$} & {97.67$_{0.05}$}\\

    \end{tabular}
    }
\vspace{-3mm}
\end{table*}

%% file: norm_ImageNet_confusion_matrix.tex
\begin{table*}[t]
    \setlength\tabcolsep{1pt} 
    \center
    \caption{HistoKT Matrix Pretrained on ImageNet (Top-1 Test Accuracy)}
    \vspace{-2mm}
	\label{tab:norm_ImageNet_confusion_matrix}
    \tiny{
    \begin{tabular}{c|P{1.6cm} P{1.6cm} P{1.6cm} P{1.6cm} P{1.6cm} P{1.6cm} P{1.6cm} P{1.6cm} P{1.6cm}}
    \hlinewd{1pt}
        \multirow{2}{*}{\backslashbox{\textbf{Source}}{\textbf{Target}}} 
        & \multirow{2}{*}{\textbf{ADP}}
        & \multirow{2}{*}{\textbf{MHIST}}
        & \multirow{2}{*}{\textbf{BACH}}
        & \multirow{2}{*}{\textbf{AJ-LYMPH}}
        & \multirow{2}{*}{\textbf{PCam}}
        & \multirow{2}{*}{\textbf{CRC}}
        & \multirow{2}{*}{\textbf{GlaS}}
        & \multirow{2}{*}{\textbf{OS}}
        & \multirow{2}{*}{\textbf{BCSS}}
        \\
        & \multirow{2}{*}{} & \multirow{2}{*}{} & \multirow{2}{*}{}& \multirow{2}{*}{} & \multirow{2}{*}{} & \multirow{2}{*}{} & \multirow{2}{*}{} & \multirow{2}{*}{}
        \\
        \hline
\textbf{ADP} & {93.03$_{0.28}$} & \cellRedBG{55.0}{76.42$_{0.21}$} & \cellRedBG{42.85}{91.63$_{0.88}$} & \cellGreenBG{100.0}{96.49$_{1.52}$} & \cellRedBG{55.0}{77.36$_{1.42}$} & \cellRedBG{41.77}{99.12$_{0.18}$} & \cellRedBG{34.0}{86.67$_{3.82}$} & \cellRedBG{55.0}{93.71$_{0.44}$} & \cellRedBG{55.0}{97.44$_{0.03}$}\\
\textbf{MHIST} & \cellGreenBG{98.57}{94.47$_{0.08}$} & {83.52$_{1.34}$} & \cellRedBG{21.17}{93.47$_{0.69}$} & \cellGreenBG{52.0}{93.86$_{1.52}$} & \cellGreenBG{74.39}{82.57$_{0.8}$} & \cellRedBG{36.47}{99.16$_{0.07}$} & \cellRedBG{13.0}{92.5$_{2.5}$} & \cellRedBG{19.38}{94.1$_{0.24}$} & \cellRedBG{12.94}{97.73$_{0.02}$}\\
\textbf{BACH} & \cellGreenBG{94.77}{94.4$_{0.06}$} & \cellGreenBG{52.59}{84.27$_{0.6}$} & {94.41$_{1.13}$} & \cellGreenBG{36.0}{92.98$_{1.52}$} & \cellGreenBG{61.42}{81.63$_{1.52}$} & \cellRedBG{23.24}{99.26$_{0.06}$} & \cellRedBG{10.0}{93.33$_{1.44}$} & \cellRedBG{34.37}{93.93$_{0.09}$} & \cellRedBG{13.47}{97.73$_{0.01}$}\\
\textbf{AJ} & \cellGreenBG{81.06}{94.15$_{0.06}$} & \cellRedBG{31.85}{80.08$_{0.77}$} & \cellRedBG{55.0}{90.6$_{0.61}$} & {92.11$_{0.0}$} & \cellGreenBG{38.56}{80.0$_{2.28}$} & \cellRedBG{41.76}{99.12$_{0.03}$} & \cellRedBG{52.0}{81.67$_{3.82}$} & \cellGreenBG{32.97}{94.32$_{0.06}$} & \cellRedBG{20.66}{97.68$_{0.08}$}\\
\textbf{PCam} & \cellGreenBG{95.18}{94.41$_{0.06}$} & \cellRedBG{19.09}{82.09$_{1.95}$} & \cellRedBG{50.73}{90.96$_{0.93}$} & \cellGreenBG{36.0}{92.98$_{1.52}$} & {78.67$_{2.57}$} & \cellRedBG{23.24}{99.26$_{0.06}$} & \cellRedBG{16.0}{91.67$_{1.44}$} & \cellRedBG{19.38}{94.1$_{0.18}$} & \cellRedBG{29.44}{97.62$_{0.09}$}\\
\textbf{CRC} & \cellGreenBG{93.62}{94.38$_{0.03}$} & \cellRedBG{22.33}{81.58$_{2.16}$} & \cellGreenBG{100.0}{94.61$_{0.82}$} & \cellRedBG{55.0}{89.47$_{9.12}$} & \cellRedBG{23.8}{78.27$_{1.37}$} & {99.35$_{0.09}$} & \cellRedBG{19.0}{90.83$_{2.89}$} & \cellGreenBG{100.0}{94.94$_{0.78}$} & \cellRedBG{44.61}{97.51$_{0.03}$}\\
\textbf{GlaS} & \cellGreenBG{100.0}{94.5$_{0.03}$} & \cellGreenBG{100.0}{85.36$_{1.85}$} & \cellRedBG{20.18}{93.55$_{0.17}$} & \cellGreenBG{84.0}{95.61$_{1.52}$} & \cellGreenBG{100.0}{84.4$_{1.81}$} & \cellRedBG{55.0}{99.03$_{0.2}$} & {93.33$_{2.89}$} & \cellGreenBG{43.78}{94.42$_{0.45}$} & \cellRedBG{11.87}{97.74$_{0.1}$}\\
\textbf{OS} & \cellGreenBG{79.54}{94.12$_{0.07}$} & \cellRedBG{25.79}{81.03$_{0.36}$} & \cellRedBG{21.5}{93.44$_{0.13}$} & \cellRedBG{10.0}{92.11$_{0.0}$} & \cellGreenBG{22.66}{78.86$_{2.88}$} & \cellRedBG{49.7}{99.07$_{0.13}$} & \cellRedBG{55.0}{80.83$_{11.81}$} & {94.2$_{0.53}$} & \cellRedBG{42.48}{97.52$_{0.1}$}\\
\textbf{BCSS} & \cellGreenBG{77.52}{94.08$_{0.09}$} & \cellRedBG{41.15}{78.61$_{1.62}$} & \cellRedBG{11.97}{94.25$_{0.29}$} & \cellGreenBG{68.0}{94.74$_{2.63}$} & \cellRedBG{27.32}{78.16$_{1.08}$} & \cellRedBG{12.65}{99.33$_{0.03}$} & \cellGreenBG{100.0}{94.17$_{2.89}$} & \cellRedBG{11.87}{94.18$_{0.42}$} & {97.75$_{0.05}$}\\

    \end{tabular}
    }
\vspace{-5mm}
\end{table*}

%% file: deepVSfineTable.tex
\newcolumntype{P}[1]{>{\centering\arraybackslash}p{#1}}
\newcolumntype{M}[1]{>{\centering\arraybackslash}m{#1}}
\begin{table*}[b]
	\vspace{-3mm}
    \setlength\tabcolsep{1pt} 
    \center
    \caption{Deep-tuning vs Fine-tuning Using ADP Pretrained Weights}
    \vspace{-2mm}
	\label{tab:deepVSfine}
    \scriptsize{
    \begin{tabular}{c|P{1.6cm}|P{1.6cm}|P{1.6cm}|P{1.6cm}|P{1.6cm}|P{1.6cm}|P{1.6cm}|P{1.6cm}}
    \hlinewd{1pt}
        \multirow{2}{*}{\backslashbox{\textbf{Tuning}}{\textbf{Applied}}} 
        & \multirow{2}{*}{\textbf{MHIST}}
        & \multirow{2}{*}{\textbf{BACH}}
        & \multirow{2}{*}{\textbf{AJ-LYMPH}}
        & \multirow{2}{*}{\textbf{PCam}}
        & \multirow{2}{*}{\textbf{CRC}}
        & \multirow{2}{*}{\textbf{GlaS}}
        & \multirow{2}{*}{\textbf{Osteosarcoma}}
        & \multirow{2}{*}{\textbf{BCSS}}
        \\
        & \multirow{2}{*}{} & \multirow{2}{*}{}& \multirow{2}{*}{} & \multirow{2}{*}{} & \multirow{2}{*}{} & \multirow{2}{*}{} & \multirow{2}{*}{}
        \\
        \hline
    \textbf{Deep-tuning} & \textcolor{ForestGreen}{78.03$_{2.20}$} & \textcolor{ForestGreen}{93.63$_{0.46}$} & \textcolor{ForestGreen}{95.61$_{4.02}$} & \textcolor{BurntOrange}{76.42$_{1.16}$} & \textcolor{ForestGreen}{99.16$_{0.07}$} & \textcolor{ForestGreen}{86.67$_{5.20}$} & \textcolor{ForestGreen}{94.52$_{0.15}$} & \textcolor{ForestGreen}{97.57$_{0.04}$}\\
    
    \textbf{Fine-tuning} & {69.53$_{1.75}$} & {65.30$_{0.22}$} & {71.93$_{5.48}$} & \textcolor{ForestGreen}{77.04$_{0.18}$} & {80.92$_{0.82}$} & {80.83$_{3.82}$} & {82.30$_{0.37}$} & {87.75$_{0.07}$}\\

    \end{tabular}
    }
\vspace{-5mm}
\end{table*}

%% file: Results.tex
\graphicspath{ {./images/main/} }

\vspace{-3mm}
\section{Results}
\vspace{-2mm}

\textbf{Single Stage Transfer Results.}
Tab.~\ref{tab:confusion_matrix} shows the results with a single stage transfer learning process. The models with the highest top-1 test accuracies, shown on the main diagonal, are used to deep tune on the applied dataset. If a deep tuned model outperforms the baseline, the result is highlighted green, and if it underperforms compared to the baseline, the result is highlighted in red.

From the results, noticeable improvement from the baseline is seen when large datasets are used as a source domain: ADP, CRC, OS, and BCSS all display the ability to transfer knowledge to other, usually smaller, datasets. Training on smaller datasets like GlaS or AJ-Lymph consistently decreased performance compared to the baseline. Of the large source domains, ADP performs poorly in tasks focused on cancer detection, such as MHIST and PCam, likely due to these task being out of domain. Note that ADP is focused only on healthy tissue, while CRC, OS, and BCSS all have diseased classes. These results show that proper choice of a source domain can affect performance, and consideration of what classes in the source domain are shared with the target domain is vital.

\textbf{Two Stage Transfer Results.}
Tab.~\ref{tab:norm_ImageNet_confusion_matrix} shows the top-1 test accuracy of all cross transferred datasets with a two-stage deep tuning process. Pretrained ImageNet weights are deep tuned using each dataset in the trained column, shown along the main diagonal, and these models are then deep tuned again with the applied dataset. 

ImageNet pretraining improves overall performance for most datasets, and produces a higher peak performance in all datasets except AJ-Lymph, which is still within margin of error. Interestingly, datasets that negatively impacted most results in single stage transfer learning, such as GlaS, BACH, and AJ-Lymph, had many more positive interactions. We posit that this is due to ImageNet pretraining providing necessary low level features that were hard for models to learn from random initialization, due to the small size of the source domain dataset. Datasets that are multi-class and are initially hard to learn (low baseline top-1 accuracy) appear to benefit most from a transfer learning process, as shown in datasets such as ADP and AJ-Lymph. 

\new{The above results describe a preliminary analysis of how much knowledge is transferable between histopathological datasets. An increased accuracy could be obtained with an experimental search of the optimal neural architecture and hyper\-parameters for each case. However, such analysis would be outside the scope of the paper.}

\input{t-sne}
\input{gradCAM}

\textbf{t-SNE and Grad-CAM Analysis.}
We chose to show t-SNE results to visualize changes to latent space representation, along with Grad-CAM to show an intuitive view on how a deep-tuned model classifies images. In Fig.~\ref{fig:tsneFig}, t-SNE plots for training and test sets for the BACH dataset are shown. Deep-tuning on the best performing dataset, CRC, we see that the representation for the test set becomes more disentangled for both single stage and two stage ImageNet training procedures. 

Using Grad-CAM, visualizations of the baseline model and transfer learning models are shown in Fig.~\ref{fig:gradCAM}. 
During generation of Grad-CAM visualizations, both augmentation smoothing and eigen smoothing were used. All activations shown are of the model output prediction, which matched with ground truth for all classes. According to expert pathologists, the ADP source domain model initialized with ImageNet weights and tuned on BACH had the best correlating activation heat map with ground truth for Benign, Invasive, and Normal classes. In contrast, the BACH model initialized with ImageNet weights (baseline model) had poor heatmap correlation with ground truth. However, both the ADP source domain model and BACH baseline model yielded the same, correct predictions for Benign, Invasive, and Normal classes. 


\begin{figure}[t]
    \centering  
    \includegraphics[width=0.8\linewidth]{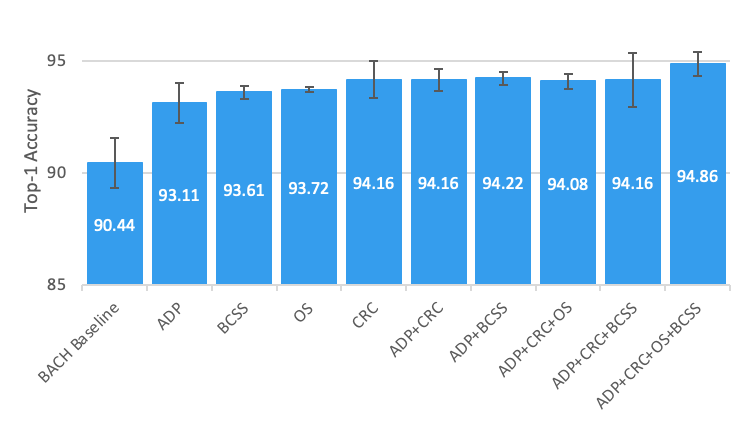}
    \vspace{-3mm}
    \caption{Different combinations of the top performing models on the BACH dataset. The models with multiple dataset labels use weight distillation for combining weights.}
    \label{fig:Weight_distillation}
    \vspace{-5mm}
\end{figure}

\textbf{Weight Distillation.}
Fig.~\ref{fig:Weight_distillation} displays top-1 accuracies of top performing models evaluated on the BACH dataset. All single dataset results, other than the baseline, are first trained on the specified dataset, and deep tuned on the BACH dataset. For the models created by weight distillation, we first apply our weight distillation workflow to combine two or more baseline weights (no tuning), and then deep tune the resulting combined model on BACH. T-SNE and Grad-CAM visualizations for weight distilled models can be found in the supplementary material.

Here, all models outperformed the BACH baseline, and all weight distilled models perform better than or equal to the one stage deep-tuning models, other than ADP + CRC + OS. We posit that this effect is due to different CPath datasets offering knowledge from different domains, shown through our Grad-CAM analysis where models trained on different source domains learned different approaches to the same task. This enables our weight distillation model to outperform even the highest performing ImageNet tuned model, demonstrating that CPath datasets hold valuable domain specific knowledge that cannot be seen in natural image datasets.

%% file: t-sne.tex
\graphicspath{ {./images/} }

\newcommand{\widthtsne}{.1\textwidth}

\begin{figure}[t]
	\centering
\begin{subfigure}{\widthtsne}
  \centering
  \fbox{\includegraphics[width=0.8\linewidth]{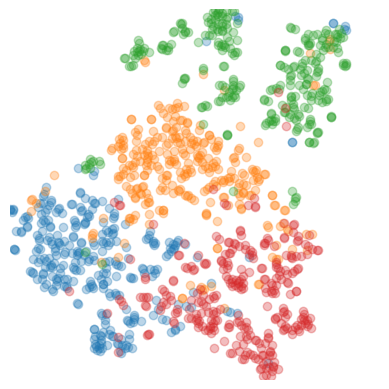}}
  \vspace{-1mm}
  \caption{}
  \label{fig:sfig2}
\end{subfigure}%
\begin{subfigure}{\widthtsne}
  \centering
  \fbox{\includegraphics[width=0.8\linewidth]{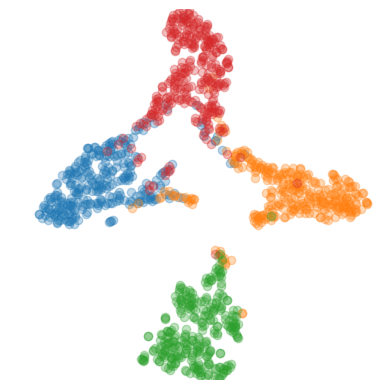}}
  \vspace{-1mm}
  \caption{}
  \label{fig:sfig4}
\end{subfigure}%
\begin{subfigure}{\widthtsne}
  \centering
  \fbox{\includegraphics[width=0.8\linewidth]{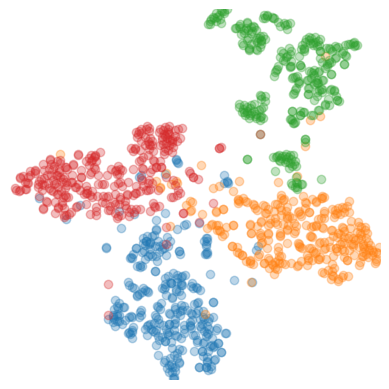}}
  \vspace{-1mm}
  \caption{}
  \label{fig:sfig7}
\end{subfigure}%
\begin{subfigure}{\widthtsne}
  \centering
  \fbox{\includegraphics[width=0.8\linewidth]{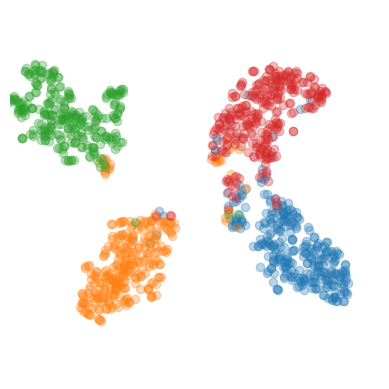}}
  \vspace{-1mm}
  \caption{}
  \label{fig:sfig9}
\end{subfigure}%
\begin{subfigure}{.07\textwidth}
  \centering
  \includegraphics[width=\linewidth]{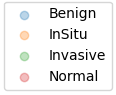}
  \label{fig:sfig10}
\end{subfigure}
\vspace{-2mm}
\caption{
	t-SNE plots of test sets: (a) BACH baseline; (b) BACH deep-tuned on CRC; (c) BACH deep-tuned on ImageNet; (d) BACH deep-tuned on CRC deep-tuned on ImageNet. 
}
\label{fig:tsneFig}
\vspace{-3mm}
\end{figure}

%% file: gradCAM.tex
\graphicspath{ {./images/} }

\begin{figure}[t]
	\scriptsize
\centering
\resizebox{\linewidth}{!}{
\setlength{\tabcolsep}{2pt}
\begin{tabular}{m{0.25\linewidth} m{0.15\linewidth} m{0.15\linewidth} m{0.15\linewidth} m{0.15\linewidth}}
    \centering{Labels} & \centering{Benign} & \centering{In Situ} & \centering{Invasive} & \multicolumn{1}{c}{Normal} \\
    Original & 
     \includegraphics[width=\linewidth]{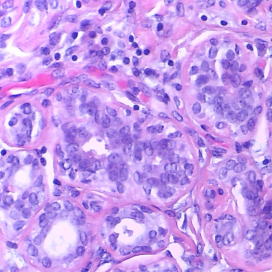} &
		\includegraphics[width=\linewidth]{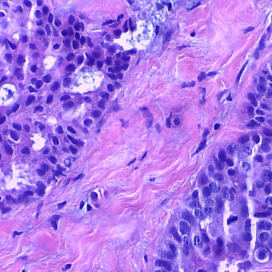} &
		\includegraphics[width=\linewidth]{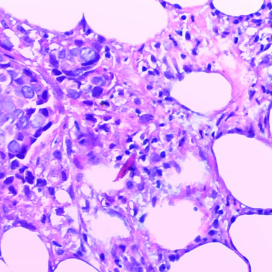} &
     \includegraphics[width=\linewidth]{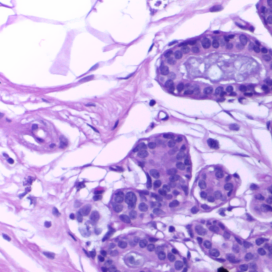} \\
    BACH Baseline & 
    \includegraphics[width=\linewidth]{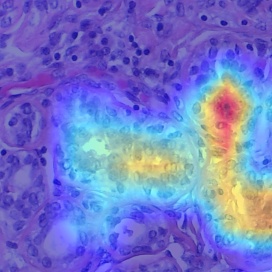} &
    \includegraphics[width=\linewidth]{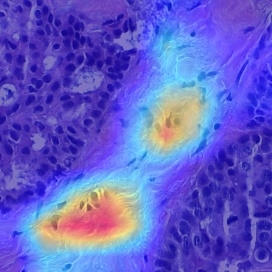} &
    \includegraphics[width=\linewidth]{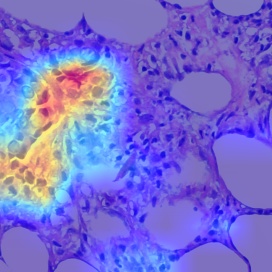} &
    \includegraphics[width=\linewidth]{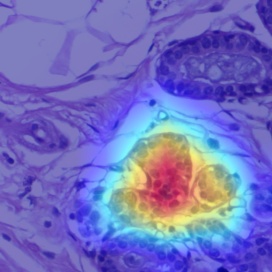} \\
    ADP Pretrained,\par BACH deep-tuned &
    \includegraphics[width=\linewidth]{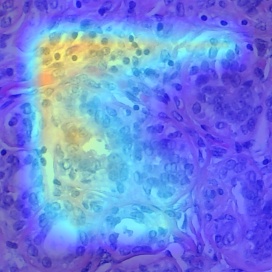} & 
    \includegraphics[width=\linewidth]{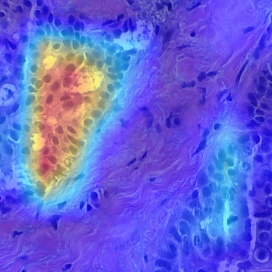} &
    \includegraphics[width=\linewidth]{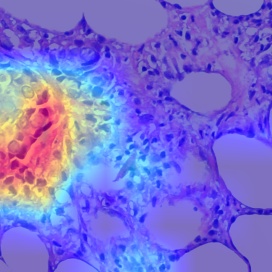} &
    \includegraphics[width=\linewidth]{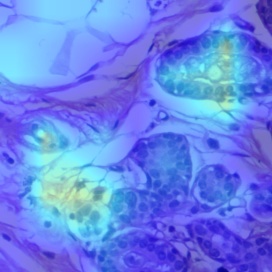} \\
    ImageNet + BACH,\par Baseline &
    \includegraphics[width=\linewidth]{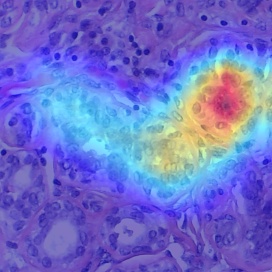} & 
    \includegraphics[width=\linewidth]{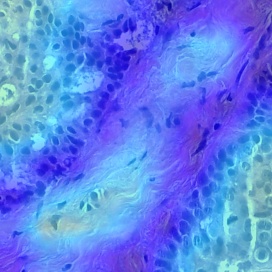} & 
    \includegraphics[width=\linewidth]{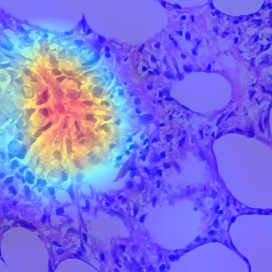} &
    \includegraphics[width=\linewidth]{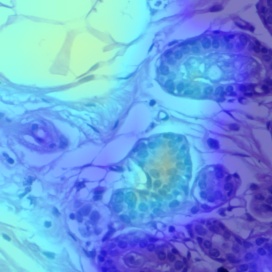} \\
    ImageNet + ADP,\par BACH deep-tuned &
    \includegraphics[width=\linewidth]{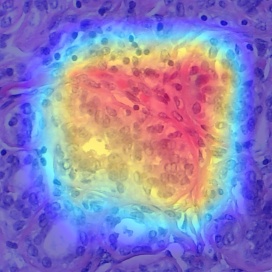} & 
    \includegraphics[width=\linewidth]{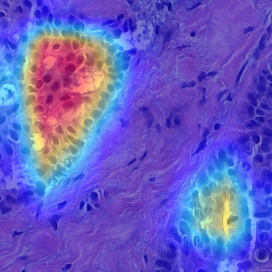} & 
    \includegraphics[width=\linewidth]{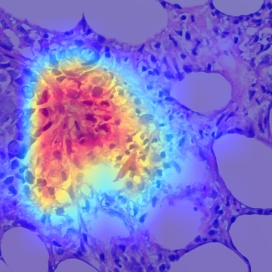} &
    \includegraphics[width=\linewidth]{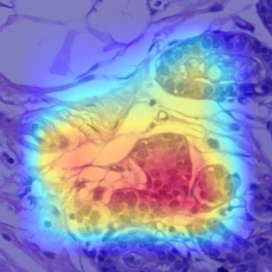} \\
\end{tabular}
}
\vspace{-3mm}
\caption{Grad-CAM images for the BACH dataset.}
\label{fig:gradCAM}
\vspace{-6mm}
\end{figure}

%% file: conclusion.tex
\vspace{-4mm}
\section{Conclusion}
\vspace{-3mm}

In this work, we proposed and tested a cross domain knowledge transfer pipeline consisting of dataset standardization, data augmentation, and training procedures over nine histopathological datasets. To assess transferred knowledge, we conducted experiments comparing source domain viability for each of the nine datasets and two stage transfer viability using ImageNet pretrained weights. To demonstrate the validity of our transfer learning framework and to visualize the learned knowledge from one dataset to another, we use t-SNE and Grad-CAM to show the change in latent space representation and class activations, respectively. Additionally, we apply weight distillation to top performing models to aggregate knowledge across datasets. We find that knowledge is transferred between histopathological datasets, and that hard to learn, multi-class datasets benefit most from transfer learning. Datasets that share a common organ class or common tasks tend to also share knowledge more effectively, especially when the constraint of learning low level features, governed by dataset size, is removed through ImageNet pretraining. These effects are also displayed through the t-SNE and Grad-CAM analysis, with more disentangled latent representations and more meaningful class activations, respectively. Weight distillation harnesses the different learned approaches by models trained on different source domains, allowing combined models to reach higher than ImageNet pretraining accuracies, with much less computational cost compared to training on ImageNet. We present these finding in an effort to push for a more data-driven approach to transfer learning in CPath, and to create a future where CPath knowledge can be shared between any number of datasets.